\documentclass[aps,prstab,preprint,groupedaddress]{revtex4}

\usepackage{graphicx}

\begin{document}

\preprint{SLAC-PUB-9227}
\preprint{KEK-Preprint 2002-26}

\title{Intrabeam scattering analysis of
measurements\\ at KEK's ATF damping ring}


\author{K.L.F. Bane}
\thanks{Work supported by the Department
of Energy, contract DE-AC03-76SF00515}
\affiliation{Stanford Linear Accelerator Center,\\
Stanford University, Stanford, CA 94309}
\author{H. Hayano, K. Kubo, T. Naito, T. Okugi, J. Urakawa}
\affiliation{High Energy Accelerator Research Organization (KEK),\\
1-1 Oho, Tsukuba, Ibaraki, Japan}


\date{\today}

\begin{abstract}

We derive a simple relation for estimating the relative emittance
growth in $x$ and $y$ due to intrabeam scattering (IBS) in electron storage
rings.
We show that IBS calculations for the ATF
damping ring, when using the formalism of Bjorken-Mtingwa,
a modified formalism of Piwinski (where $\eta^2/\beta$
has been replaced by ${\cal H}$), or a simple high-energy
approximate formula all give results that agree well.
Comparing theory,
including the effect of potential well bunch lengthening,
with a complete set of ATF steady-state
beam size {\it vs.} current measurements
we find reasonably good
agreement for energy spread and horizontal emittance.
The measured vertical emittance, however,
is larger than theory in both offset (zero
current emittance) and slope (emittance change with current).
The slope error indicates measurement error and/or
additional current-dependent physics at the ATF;
the offset error, that the assumed Coulomb log is
correct to within a factor of 1.75.

\end{abstract}


\maketitle

\section{INTRODUCTION}

In future e$^+$e$^-$ linear colliders, such as the JLC/NLC\cite{JLC,NLC}, damping
rings are needed to generate beams of intense bunches with
low emittances.
The Accelerator Test Facility
(ATF)\cite{Hinode:95} at KEK is a prototype of such damping
rings.
One of its main goals, and one that
has been achieved, was the demonstration of extremely low vertical
emittances\cite{ATF:02,Sakai:02}. At the low ATF emittances, however, it is
found that intrabeam scattering (IBS) is a strong effect, and
one that needs to be understood.

Intrabeam scattering is an effect that depends on the ring
lattice---including the errors---and on all dimensions of the
beam, including the energy spread. At the ATF all these dimensions
can be measured; unique to the ATF is that the beam energy spread,
an especially important parameter in IBS theory, can be measured
to an accuracy of a few percent. In April 2000 the single bunch
energy spread, bunch length, and horizontal and vertical
emittances were all measured as functions of current over a short
period of
time\cite{Urakawa:00,Bane:01}. 
The short period of time was important to ensure that
the machine conditions remained unchanged;
the bunch length measurement was important since
potential well bunch lengthening is significant
at the ATF\cite{Bane:01}.
The question that we attempt to answer here is,
Are these measurement results in accord with IBS theory?

Intrabeam scattering theory was first developed for accelerators by
Piwinski\cite{Piwinski:74}, a result that was extended by
Martini\cite{Martini:84}, to give a formulation that we call here
the standard Piwinski~(P) method\cite{Piwinski:99}; this was
followed by the equally detailed Bjorken and Mtingwa (B-M)
result\cite{Bjorken:83}. Both approaches solve the local,
two-particle Coulomb scattering problem for (six-dimensional)
Gaussian, uncoupled beams, but the two results appear to be
different; of the two, the B-M result is thought to be the more
general\cite{Piwinski:p}. Other simpler, more approximate
formulations developed over the years are ones due to
Parzen\cite{Parzen:87}, Le Duff\cite{LeDuff:89},
Raubenheimer\cite{Raubenheimer:91}, and Wei\cite{Wei:93}. Recent
reports on IBS theory include one by Kubo and Oide, who adapt an
intermediate result from Bjorken-Mtingwa's paper to find the
solution for cases of arbitrary coupling\cite{Kubo:01b}, a method
that is now used in the optics computer program SAD\cite{Oide:s};
and one by Venturini that solves for IBS in the presence of a
strong ring impedance\cite{Venturini:01}.

Intrabeam scattering measurements have been performed primarily on
hadronic\cite{Conte:85,Evans:86,Bhat:99,Zorzano:00} and heavy ion
machines\cite{Rao:00,RHIC:01}, where the effect tends to be more
pronounced, though measurement reports on low emittance electron
rings can also be found\cite{KimC:98,Steier:01}. Typical of such
reports, however, is that although good agreement may be found in
some beam dimension(s), the set of measurements and/or agreement
is not complete ({\it e.g.} in Ref.~\cite{Conte:85} growth rates
agree reasonably well in the longitudinal and horizontal
directions, but completely disagree in the vertical). Note that
one advantage of studying IBS using electron machines is that it
can be done by measuring steady-state beam sizes.

In this report we briefly describe
intrabeam scattering formulations, apply and
compare them for ATF parameters, and finally compare calculations with the
full set of data of April 2000. For more details on the
hardware and such measurements at the ATF, the reader is referred
to Ref.~\cite{ATF:02,Sakai:02}.

\section{IBS CALCULATIONS}

We begin by describing the method of calculating the effect
of IBS in a storage ring.
Let us first assume that there is no $x$-$y$ coupling.

Let us consider the IBS growth rates in energy $p$, in the
horizontal $x$, and in the vertical $y$ to be defined as
\begin{equation}
{1\over T_p}={1\over\sigma_p}{d\sigma_p\over dt}\ ,\quad {1\over
T_x}={1\over\epsilon_x^{1/2}}{d\epsilon_x^{1/2}\over dt}\ ,\quad
{1\over T_y}={1\over\epsilon_y^{1/2}}{d\epsilon_y^{1/2}\over dt}\
.\label{growth_rates_eq}
\end{equation}
Here $\sigma_p$ is the rms (relative) energy spread,
 $\epsilon_x$ the horizontal emittance, and
$\epsilon_y$ the vertical emittance.
In general, the growth rates are given in both P and B-M theories
in the form:
\begin{equation}
{1\over T_i}=\left< f_i\right> \label{tinv_eq}
\end{equation}
where subscript $i$ stands for $p$, $x$, or $y$.
The functions
$f_i$ are integrals that depend on beam parameters, such as energy
and phase space density, and lattice properties, including
dispersion;
the brackets $\langle\rangle$ mean that the quantity is averaged
over the ring.
In this report we will primarily use the $f_i$
of the B-M formulation\footnote{We believe
that the right hand side of Eq.~4.17 in B-M (with $\sigma_\eta$
equal to our $\sqrt{2}\sigma_p$) should be divided by $\sqrt{2}$,
in agreement with the recent derivation of
Ref.~\cite{Venturini:01}. Also, vertical dispersion,
though not originally in B-M, can be
added in the same manner as horizontal dispersion. }.


From the $1/T_i$ we obtain the steady-state properties for
machines with radiation damping:
\begin{equation}
\epsilon_x={\epsilon_{x0}\over 1-\tau_x/T_x}\ ,\
\epsilon_y={\epsilon_{y0}\over 1-\tau_y/T_y}\ ,\
\sigma^2_{p}={\sigma^2_{p0}\over 1-\tau_{p}/T_{p}}\ ,
\label{eqiterate}
\end{equation}
where subscript 0 represents the beam property due to synchrotron
radiation alone, {\it i.e.} in the absence of IBS, and the
$\tau_i$ are synchrotron radiation damping times. These are 3
coupled equations since all 3 IBS rise times depend on
$\epsilon_x$, $\epsilon_y$, and $\sigma_p$.



The way of solving Eqs.~\ref{eqiterate} that we employ is to convert them into
3 coupled differential equations, such as is done in {\it e.g.}
Ref.~\cite{KimC:97}, and solve for the asymptotic values. For
example, the equation for $\epsilon_y$ becomes
\begin{equation}
{d\epsilon_y\over dt}= -{2(\epsilon_y-\epsilon_{y0})\over \tau_y}
+{2\epsilon_y\over T_y}\quad,\label{epsydif_eq}
\end{equation}
and there are corresponding
equations for $\epsilon_x$ and $\sigma_p^2$.

Before solving these equations one needs to know the source
of the vertical emittance at zero
current. We consider 3 possible sources: (i)~vertical
dispersion due to vertical orbit errors, (ii)~(weak) $x$-$y$ coupling
due to such things as rolled quads, etc, and (iii)~a combination
of the two.
If the vertical emittance at zero
current is due mainly to vertical dispersion,
then\cite{Raubenheimer:91}
\begin{equation}
\epsilon_{y0}\approx {\cal J}_\epsilon\langle{\cal
H}_y\rangle\sigma_{p0}^2\quad, \label{Tor0_eq}
\end{equation}
with ${\cal J}_\epsilon$ the energy damping partition number and
${\cal H}=[\eta^2+(\beta\eta^\prime-{1\over 2}\beta^\prime\eta)^2]/\beta$
the dispersion invariant,
with $\eta$ and $\beta$,
respectively, the lattice dispersion and beta functions. If $\epsilon_{y0}$
 is mainly due to coupling
we drop the $\epsilon_y$ differential equation and simply
let $\epsilon_y=\kappa\epsilon_x$, with $\kappa$ the coupling factor.
In case (iii)
we approximate
the solution by replacing the parameter $\epsilon_{y0}$ in
Eq.~\ref{epsydif_eq} by
the quantity [$\kappa\epsilon_x(1-\tau_y/T_y)+\epsilon_{y0d}$],
where $\epsilon_{y0d}$ is the part of $\epsilon_{y0}$
due to dispersion only.
Note that the practice---sometimes
found in the literature---of solving IBS equations assuming
no vertical errors, which tends to result in near 0 or even negative
vertical emittance growth, may describe a state that is unrealistic
and unachievable.
Note also that in case~(i)
once the vertical
orbit---and therefore $\langle{\cal H}_y\rangle$---is set,
$\epsilon_{y0}$ is no longer a free parameter.

In addition, note that:

\begin{itemize}

\item
A fourth
equation in our system, the relation between bunch length $\sigma_s$ and
$\sigma_p$, is also implied; generally this is taken to be the
nominal (zero current) relation.
In the ATF strong potential well bunch lengthening,
though no microwave instability, is found at the highest single bunch
currents\cite{Bane:01}.
In our comparisons with ATF measurements
we approximate this effect
by adding a
multiplicative factor $f_{pw}(I)$ [$I$ is current], obtained from
measurements, to the equation relating $\sigma_s$ to $\sigma_p$.
(Note that potential well bunch lengthening
also changes the longitudinal bunch shape, a less important effect
that we will ignore.)

\item The B-M results include a so-called Coulomb log factor,
$\ln(b_{max}/b_{min})$, with $b_{max}$, $b_{min}$ maximum, minimum
impact parameters, quantities which are not well defined.
For round beams it seems that $b_{max}$ should
be taken as the beam size
\cite{Farouki:94}.
For bi-Gaussian beams it is not clear
what the right choice is. Normally $b_{max}$ is taken to be the
vertical beams size, though sometimes the horizontal beam size
is chosen\cite{Wei:01}.
We take $b_{max}=\sigma_y$; $b_{min}=r_0 c^2/\langle v_x^2\rangle$
$=r_0\beta_x/(\gamma^2\epsilon_x)$, with $r_0$ the classical
electron radius ($=2.82\times10^{-15}$~m),
$v_x$ the transverse velocity in the rest frame,
and $\gamma$ the Lorentz
energy factor. For the ATF, the Coulomb log, $({\rm log})=16.0$.

\item The IBS bunch distributions
are not Gaussian, and tail particles can be overemphasized in
these solutions. We are interested in core sizes, which we
estimate by eliminating interactions with collision rates less
than the synchrotron radiation damping rate\cite{Raubenheimer:94}.
We can approximate this in the Coulomb log term by letting $\pi
b_{min}^2\langle |v_x|\rangle\langle n\rangle$ equal
the synchrotron damping rate in the rest frame, with $n$ the
particle density in the rest frame\cite{Kubo:01b}; or
$b_{min}=\sqrt{4\pi\sigma_x\sigma_y\sigma_z\gamma/[Nc\tau]}
(\beta_x/\epsilon_x)^{1/4}$, with $N$ the bunch population. For
the ATF with this cut, $({\rm log})=10.0$.

\end{itemize}

\subsection{High Energy Approximation}

For both the P and the B-M methods solving
for the IBS growth rates is time consuming,
involving, at each iteration step, a numerical integration
at every lattice element.
A quicker-to-calculate, high energy approximation,
one valid
in normal storage ring lattices,
can be derived from the B-M formalism\cite{Bane:02}:
\begin{eqnarray}
{1\over T_p}& \approx & {r_0^2 cN({\rm log})\over
16\gamma^3\epsilon_x^{3/4}\epsilon_y^{3/4}\sigma_s\sigma_p^3}
\left<\sigma_H\,
g(a/b)\,\left({\beta_x\beta_y}\right)^{-1/4}\right>
 \nonumber\\
{1\over T_{x,y}}& \approx & {\sigma_p^2\langle{\cal H}_{x,y}
\rangle\over\epsilon_{x,y}}{1\over T_p}
\quad,\label{Tor_eq}
\end{eqnarray}
with
\begin{equation}
{1\over\sigma_H^2}= {1\over\sigma_p^2} +
{{\cal H}_x\over\epsilon_x} +
{{\cal H}_y\over\epsilon_y}\quad,
\end{equation}
\begin{equation}
a={\sigma_H\over\gamma}
\sqrt{\beta_x\over\epsilon_x}\quad,\quad\quad
b={\sigma_H\over\gamma}\sqrt{\beta_y\over\epsilon_y}\quad.
\label{ab_rev_eq}
\end{equation}
The requirement on high energy is that $a$,$b\ll1$;
if it is satisfied then the beam momentum in the longitudinal
plane is much less than in the transverse planes.
For flat beams $a/b$
is less than 1. In the ATF, for example, when
$\epsilon_y/\epsilon_x\sim0.01$,
$a\sim0.01$, $b\sim0.1$, and $a/b\sim0.1$. The function $g$,
related to the elliptic integral,
can be well approximated by
\begin{equation}
g(\alpha)\approx\alpha^{(0.021-0.044\ln\alpha)}\quad\quad
[{\rm for}\ 0.01<\alpha<1]\ ;\label{gmain_eq}
\end{equation}
to obtain $g$ for $\alpha>1$, note that $g(\alpha)=g(1/\alpha)$.

Note that
Parzen's high energy formula is a similar,
though more approximate, result to that given here\cite{Parzen:87};
and Raubenheimer's approximation
is formulas similar, though less accurate, than the first
and identical to the 2nd and 3rd of Eqs.~\ref{Tor_eq}\cite{Raubenheimer:91}.
Note that Eqs.~\ref{Tor_eq} assume that $\epsilon_{y0}$ is due
mainly to vertical dispersion; if it is due mainly to
$x$-$y$ coupling we
let ${\cal H}_y=0$, drop the $1/T_y$ equation, and simply let
$\epsilon_y=\kappa\epsilon_x$.
Finally, note that these equations still need to be iterated, as described before, to
find the steady-state solutions.

\subsection{Emittance Growth Theorem}\label{theorem_sec}


Following an argument in Ref.~\cite{Raubenheimer:91} we can obtain
a relation between the expected vertical and horizontal emittance
growth due to IBS in the presence of random vertical dispersion:
We begin by noting
that the beam momentum in the longitudinal plane is much less than in
the transverse planes. Therefore, IBS will first heat the
longitudinal plane; this, in turn, increases the transverse
emittances through dispersion (through ${\cal H}$,
as can be seen in the 2nd and 3rd of Eqs.~\ref{Tor_eq}), like
synchrotron radiation (SR) does. One difference between IBS and SR
is that IBS increases the emittance everywhere, and SR only in
bends. We can write
\begin{equation}
{\epsilon_{y0}\over\epsilon_{x0}}\approx {{\cal J}_x\langle{\cal
H}_y\rangle_{b} \over{\cal J}_y\langle{\cal H}_x\rangle_{b}} \quad
,\quad
{\epsilon_{y}-\epsilon_{y0}\over\epsilon_{x}-\epsilon_{x0}}\approx
{{\cal J}_x\langle{\cal H}_y\rangle\over{\cal J}_y\langle{\cal
H}_x\rangle}\ ,
\end{equation}
where ${\cal J}_{x,y}$ are damping partition numbers, and $\langle\rangle_b$
 means averaging is only done over the bends. For
vertical dispersion due to errors we expect $\langle{\cal
H}_y\rangle_{b}\approx\langle{\cal H}_y\rangle$. Therefore,
\begin{equation}
r_\epsilon\equiv {(\epsilon_{y}-\epsilon_{y0})/\epsilon_{y0}
\over(\epsilon_{x}-\epsilon_{x0})/\epsilon_{x0}}\approx
{\langle{\cal H}_x\rangle_{b}\over\langle{\cal H}_x\rangle}
\quad,\label{Kubo_eq}
\end{equation}
which, for the ATF is 1.6. If, however, there is only $x$-$y$
coupling, $r_\epsilon=1$; if there is both vertical dispersion and
coupling, $r_\epsilon$ will be between $\langle{\cal
H}_x\rangle_{b}/\langle{\cal H}_x\rangle$ and 1.

\subsection{Numerical Comparison}

Let us compare the results of the methods P, B-M, and Eq.~\ref{Tor_eq}
 when
applied to the ATF beam parameters and lattice, with vertical
dispersion and no $x$-$y$ coupling. We take as parameters
those given in Table~I, and, for this comparison, let
$f_{pw}=1$. In addition we have
${\cal J}_\epsilon=1.4$, $\langle\beta_x\rangle=3.9$~m,
$\langle\beta_y\rangle=4.5$~m, $\langle \eta_x\rangle=5.2$~cm and
$\langle{\cal H}_x\rangle=2.9$~mm. To generate vertical dispersion
we randomly offset magnets by 15~$\mu$m, and then calculate the
closed orbit using SAD. For our seed we find that the rms
dispersion $(\eta_y)_{rms}=7.4$~mm, $\langle{\cal
H}_y\rangle=17$~$\mu$m, and $\epsilon_{y0}=6.9$~pm (in agreement
with Eq.~\ref{Tor0_eq}). For consistency between the methods we
here take the cut-off parameter $d=3\sigma_y$ in P to corresponds to
$({\rm log})=\ln{[d\sigma_H^2/(4r_0a^2)]}=16$ in B-M.

\begin{table}[htb]
\caption{
Typical ATF parameters in single bunch mode.
}
\begin{ruledtabular}
\begin{tabular}{l c c l}
Circumference & $C$ & 138 & m\\
Energy & $E$ & 1.28 & GeV\\
Current & $I$ & 3.1 & mA\\
Nominal energy spread & $\sigma_{p0}$ & $5.44$ & $10^{-4}$\\
Nominal horizontal emittance & $\epsilon_{x0}$ & 1.05 & nm\\
Nominal bunch length & $\sigma_{s0}$ & 5.06\footnote{at rf voltage 300~kV} & mm\\
Longitudinal damping time & $\tau_p$ & 20.9 & ms\\
Horizontal damping time & $\tau_x$ & 18.2 & ms\\
Vertical damping time & $\tau_y$ & 29.2 & ms
\end{tabular}
\end{ruledtabular}
\end{table}

\begin{figure}[htb]
\centering
\includegraphics*[width=80mm]{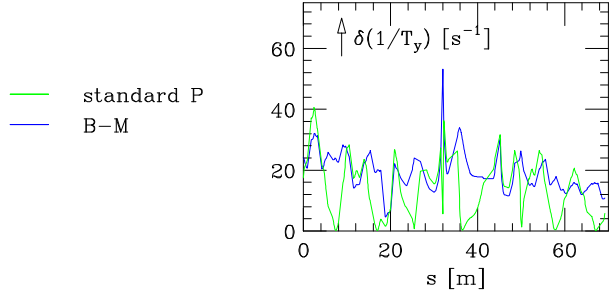}
\caption{ Vertical steady-state (local) growth rate over 1/2 the ATF
for an example with vertical dispersion due to random errors. Given are
results due to
 standard Piwinski (green) and  Bjorken-Mtingwa (blue).
 }
\label{fidebug0}
\end{figure}

Performing the calculations, but first comparing the standard
Piwinski and the B-M methods, we find that the growth rates in $p$
and $x$ agree well; the vertical rate, however, does not. In
Fig.~\ref{fidebug0} we display the {\it local} IBS growth rate in
$y$ over half the ring (the periodicity is 2), as obtained by the
two methods, and see that the P result, on average, is 25\% low.
Studying the two methods we note that a conspicuous difference
between them is their dependence on dispersion: for P the $f_i$
depend on it only through $\eta^2/\beta$; for B-M, through
$\phi=[\eta^\prime-{1\over 2}\beta^\prime\eta/\beta]$ and through
${\cal H}$. Let us replace $\eta^2/\beta$ in P with ${\cal H}$ to
create a method that we call the {\it modified} Piwinski result.
In Ref.~\cite{Bane:02} it is shown that, in a normal storage ring
lattice, at high energies, the results of this method become equal
to those of B-M.

\begin{figure}[htb]
\centering
\includegraphics*[width=82mm]{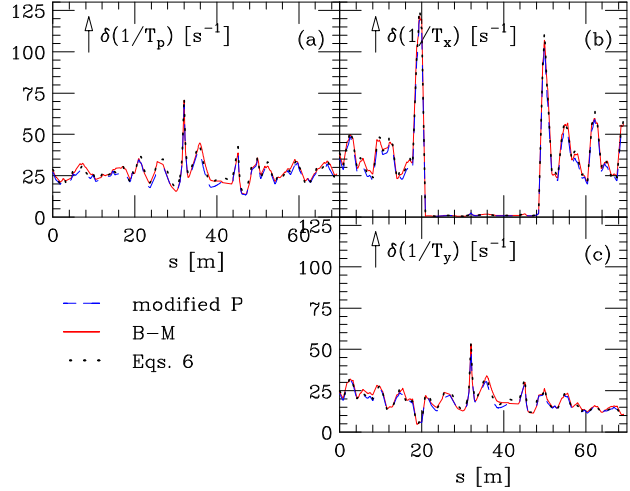}
\caption{ Steady-state (local) growth rates over 1/2 the ATF
for an example with vertical dispersion due to random errors. Given are
results due to
 modified Piwinski, Bjorken-Mtingwa, and Eqs.~\ref{Tor_eq}.
 }
\label{fidebug}
\end{figure}

Comparing with this method we find that, indeed, the three growth
rates now agree reasonably well with the B-M result.
Fig.~\ref{fidebug} displays the 3 local growth rates as obtained by
the modified P and B-M methods. The $1/T_i$, the average values of
these functions, are given in Table~II. We note that the P results
are all slightly low, by 4.5\%. The B-M method gives:
$\sigma_p/\sigma_{p0}=1.52$, $\epsilon_x/\epsilon_{x0}=1.90$,
$\epsilon_y/\epsilon_{y0}=2.30$. Note that for this error seed the
emittance growth ratio of Eq.~\ref{Kubo_eq} is $r_\epsilon=1.44$,
close to the 1.6 expected for the ATF lattice.

\begin{table}[htb]
\caption{
Steady-state IBS growth rates for an example including vertical dispersion
due to random errors.}
\begin{ruledtabular}
\begin{tabular}{l c c c}
Method & $1/T_p$ [s$^{-1}$] & $1/T_x$ [s$^{-1}$] & $1/T_y$ [s$^{-1}$]\\ \hline\hline
Modified Piwinski & 25.9 & 24.7  & 18.5 \\
Bjorken-Mtingwa & 27.0 & 26.0 & 19.4\\
Eqs.~\ref{Tor_eq} & 27.4 & 26.0 & 19.4\\
\end{tabular}
\end{ruledtabular}
\end{table}

Repeating the calculation using Eqs.~\ref{Tor_eq}
we find that the computing time is greatly reduced, and the
growth rates agree quite well with the B-M results (see Table~II).
The dots in Fig.~\ref{fidebug} give the local rates
corresponding to Eqs.~\ref{Tor_eq}, and we
see that even these agree quite well.

\subsection{Comparison with SAD Results}

The optics program SAD basically follows the B-M formalism,
but it does it in a form that treats the three
beam directions on equal footing.
The final results are given in terms of the normal modes of the system and
not the beta and dispersion functions of the uncoupled system
(as in our approximation).
For vertical dispersion dominated problems there is no difference in result.
In coupling dominated problems there will be a difference
that, in the case of small
$x$-$y$ coupling due to errors, we expect to be small.

\begin{figure}[htb]
\centering
\includegraphics*[width=80mm]{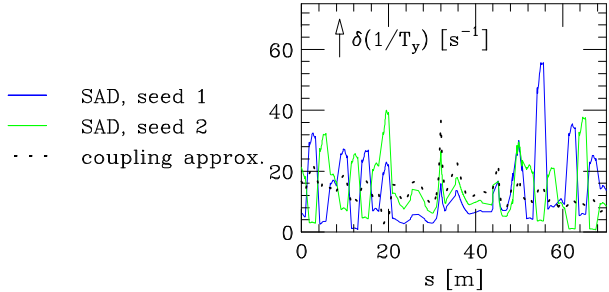}
\caption{
Vertical steady-state (local) growth rate over 1/2 the ATF
for an example with vertical dispersion
and $x$-$y$ coupling due to random errors. Given are
results obtained
by SAD (for 2 seeds; the solid curves) and by the coupling approximation
used here (the dots).
 }
\label{fisad}
\end{figure}

We consider the ATF lattice with random magnet offsets and rotations.
Other machine parameters are the same as before; again $I=3.1$~mA.
For this lattice $(\eta_y)_{rms}=7.4$~mm and
$\epsilon_{y0}/\epsilon_{x0}=1\%$. For this problem we solve IBS
using SAD (for 2 different seeds), and also our approximate
method where we include vertical dispersion (as before)
and a global
coupling parameter $\kappa=0.34\%$. We take $({\rm log})=9.1$.
Comparing steady-state {\it local} growth rates, we find
good agreement in $p$ and $x$ for all three calculations.
In $y$, however, there is a significant variation (see Fig.~\ref{fisad}).
The growth rates, the average values
of these functions, however, agree well
(see Table~III). Note that the
steady-state relative growths in ($\sigma_p$,$\epsilon_x$,$\epsilon_y$)
are
(1.38,1.56,1.64) for SAD, and
(1.38,1.62,1.61) for our approximate calculation.

\begin{table}[htb]
\caption{
Steady-state IBS growth rates for an example including vertical dispersion and
$x$-$y$ coupling due to random errors.
}
\begin{ruledtabular}
\begin{tabular}{l c c c}
Method & $1/T_p$ [s$^{-1}$] & $1/T_x$ [s$^{-1}$] & $1/T_y$ [s$^{-1}$]\\ \hline\hline
SAD, seed 1 & 22.5 & 19.6  & 13.1 \\
SAD, seed 2 & 22.3 & 19.6 & 13.5\\
Our approx. calculation & 22.9 & 21.0 & 12.9\\
\end{tabular}
\end{ruledtabular}
\end{table}

\section{COMPARISON WITH MEASUREMENT}

\subsection{Measurements}

At the ATF the energy spread and all beam sizes
can be measured. Unique at the ATF is that the
energy spread, a particularly important
parameter in IBS theory, can be obtained to a few percent accuracy.
In this measurement the beam is extracted and its size measured
on a screen in a highly dispersive region.
The bunch length is determined with a streak camera in the ring.
The emittances can be measured using 3 methods: wire monitors
in the extraction line, a laser wire in the ring, and
an interferometer in the ring. Unfortunately, for $\epsilon_y$
all 3 methods have their difficulties. For example,
the wire measurement is very sensitive to
optics errors (such as roll and dispersion) in the extraction line.
Or, the
laser wire measurement, being time consuming (taking
$\sim1$~hour per measurement),
is sensitive to drifts in machine and beam properties.

Because of the effects of IBS
the energy spread measurement (which is quick and easy to
perform)
has become a useful technique for monitoring changes in beam size.
Thus, evidence that we are truly seeing IBS at the ATF include:
(1)~when moving onto the coupling resonance,
the normally large energy spread growth with current becomes
negligibly small; (2)~if we decrease the vertical emittance using
dispersion correction, the energy spread increases.

\subsection{Comparison with Theory}

\begin{figure}[htb]
\centering
\includegraphics*[width=65mm]{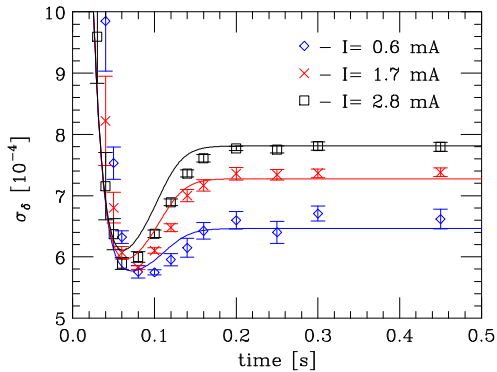}
\caption{
Measured energy spread as function of time after injection, for 3 different
currents (the plotting symbols). The curves give B-M simulations
assuming an $x$-$y$ coupling of 0.006 and no potential well distortion.
This plot is reproduced from Ref.~\cite{ATF:02}.
 }\label{sige_vs_t_eq}
\end{figure}

In Fig.~\ref{sige_vs_t_eq}, as an example, we present
the time development, after injection, of energy spread
for 3 different beam currents (the plotting symbols).
The measurement was performed by continually injecting beam
into the ATF, while varying the extraction timing.
If we take the B-M formalism, with $f_{pw}=1$, and with
$x$-$y$ coupling 0.006, and solve
the differential equations for energy spread and beam
sizes, we obtain the curves in the figure
(if we include potential well distortion the fitted coupling
becomes 0.0045).
The short time ($\lesssim0.05$~s) behavior does not agree with the data,
since the beam in reality enters the ring badly mismatched
(a region which would be difficult to simulate); in the longer time
range, however, after $\gtrsim3\tau_p$, the agreement becomes quite good.
The minimum in the curves can be explained as follows:
Initially the energy spread and beam sizes reduce due to synchrotron
radiation; when the beam volume becomes smaller than a certain
amount, the energy spread begins to increase due to IBS.
This result indicates reasonably good agreement between
measurement and theory.

\begin{figure}[htb]
\centering
\includegraphics*[width=86mm]{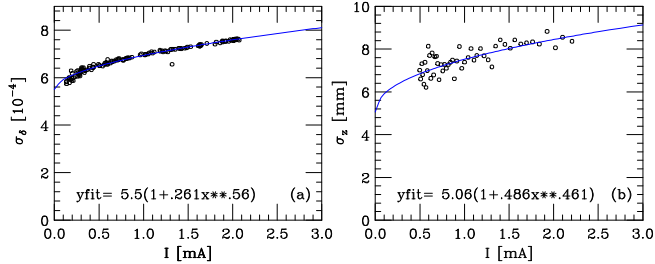}
\caption{ Measurements of steady-state energy spread~(a) and bunch length~(b),
 with $V_c=300$~kV.
 } \label{fisigpsigz}
\end{figure}

\begin{figure}[htb]
\centering
\includegraphics*[width=85mm]{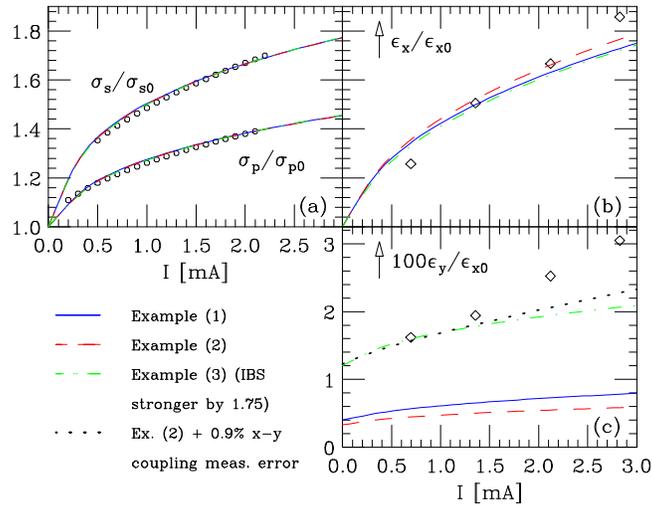}
\caption{ATF measurement data (symbols) and IBS theory fits (the
curves). The symbols in (a) give the smooth curve fits to the
measured data of Fig.~\ref{fisigpsigz}.
 } \label{fifit}
\end{figure}

To compare with theory absolutely, however, we need to measure all beam
properties with the machine in the same condition.
Such a complete series of measurements was performed
on stored beam at the ATF
over a short period of time in April 2000. The rf voltage was $V_c=300$~kV.
The energy spread
and bunch length {\it vs.} current
measurements are shown in Fig.~\ref{fisigpsigz}.
The curves in the plots are fits that
give the expected zero current result. Emittances were measured on the
wire monitors in the extraction line (the symbols in
Fig.~\ref{fifit}b-c; note that the symbols in Fig.~\ref{fifit}a
reproduce the fits to the data of Fig.~\ref{fisigpsigz}).
We see large growth also in the emittances.
Unfortunately, we have no error bars for the emittance measurements,
though we
expect the random component of errors in $y$ to be 5-10\%, and
less in $x$. Note that $\epsilon_{y0}$ appears to be about 1.0-1.2\% of
$\epsilon_{x0}$.

Let us compare B-M calculations with the data. Here we take
$f_{pw}$ as given by the measurements, and take
as Coulomb log our best estimate, $({\rm log})=10$.
Note that in the machine
the residual dispersion is typically $(\eta_y)_{rms}\sim3$~mm.
To set our one free parameter, $\epsilon_{y0}$, we adjust it
until at high current $\sigma_p$
agrees with the measurement.
In Fig.~\ref{fifit} we give examples:
\begin{enumerate}
\item Vertical dispersion only, with $(\eta_y)_{rms}=5.6$~mm
and $\epsilon_{y0}=4.0$~pm (solid);
\item Coupling dominated with $\kappa=0.33\%$ (dashes);
\item Coupling dominated with $\kappa=1.2\%$, with the
Coulomb log artificially increased
by a factor 1.75
(dotdash);
\item Same as Ex.~2 but assuming
$\epsilon_y$ measurement error, {\it i.e.} adding 0.9\% of the
measured (and splined) $\epsilon_x$ to the calculated $\epsilon_y$
(the dots).
\end{enumerate}

We see that $\sigma_p(I)$ agrees well with the measurements
for all cases, and
$\epsilon_x(I)$ agrees reasonably well. For Examples~1 and 2, however,
$\epsilon_{y0}$ is significantly lower than the measurements
seem to indicate, and the growth with current is also less.
To obtain reasonable agreement for $\epsilon_{y0}$
we need to assume that either
IBS is $\sim75\%$ stronger (in growth rates) than theory
predicts, or there is significant measurement error, equivalent
to $\sim1\%$ $x$ emittance coupling into the $y$ measurement.
Yet even with such assumptions the $\epsilon_y(I)$ dependence
does not agree.

What does the emittance growth theorem
of Sec.~\ref{theorem_sec} say about these results?
It appears that
$\epsilon_x$ grows by $\sim85\%$ by $I=3$~mA;
$\epsilon_y$ begins at about 1.0-1.2\% of $\epsilon_{x0}$, and
then grows to about 3\% of $\epsilon_{x0}$.
Therefore, the relative emittance growth ratio is
$r_\epsilon\sim2.1$--2.4, much larger than
the expected result if we are coupling dominated (1.0); and still
significantly larger than the expected
result if we are dispersion dominated (1.6),
a case that is anyway unlikely since it requires an implausibly large
$(\eta_{y})_{rms}\approx9$~mm. Thus, the emittance growth
theorem indicates that $\epsilon_y(I)$ as measured is not in agreement
with IBS theory.

\section{DISCUSSION}

Our disagreement in $\epsilon_y$ between theory and measurement
consists of
two parts, an offset part ($\epsilon_{y0}$) and a disagreement
in slope ($d\epsilon_y/dI$).
Together they indicate that we have:
error in theory,
additional physics at the ATF, and /or
error in measurement.

IBS theory is a mature theory, and the relation between longitudinal
and transverse growth rates (the 2nd and 3rd of Eqs.~\ref{Tor_eq})
is simple and intuitively easy to understand.
The main uncertainty in theory may be with the scale factor, particularly in
the Coulomb log factor for beams
with elliptical cross-section.
Yet a scale factor error can affect only the offset part of the disagreement.
Note also that even if the argument of (log) were in error
by an order of magnitude
this part of the disagreement would be changed by
only a small amount (25\%).

The disagreement in $d\epsilon_y/dI$
might be explained by the presence of additional
current-dependent physics at the ATF. We have
seen that $\sigma_p(I)$ and
$\epsilon_x(I)$ can be made to agree reasonably well between theory
and measurement;
at the same time, however, the measured $\epsilon_y(I)$ grows
much faster than predicted.
One might, therefore, suspect the presence at the ATF
of another current-dependent
effect, one that
increases the projected vertical emittance---though not the real
emittance. An example of such an effect is a $y$-$z$ tilt
of the beam induced by closed orbit distortion in the presence
of a transverse impedance\cite{Chao:83,Raubenheimer:91}.
More study needs to be done in this direction.

As mentioned before, measuring accurately
the small vertical emittances at the ATF is
difficult, and, therefore,
emittance measurement error is likely responsible
for much of the disagreement found.
We noted that a 1\% coupling measurement error in the
extraction line wire measurements can account for the
offset part of the disagreement;
the slope disagreement, however, is not easy to explain
assuming measurement error alone
(for an attempt in this direction, see {\it e.g.}
Ref.~\cite{Ross:01}).

Over the time since April 2000 the systematics of the emittance measurements
have improved, especially for the laser wire measurement.
Newer results
seem to suggest that the April 2000 measured vertical emittance
may have been too large\cite{ATF:02,Sakai:02}.
For the near future we urge
that the effort to obtain reliable emittance measurements
at the ATF be continued.
In addition, experiments to study the
possible existence of
other current-dependent effects should
also be performed.
Ultimately, one goal should be to test the accuracy of theoretical IBS growth
rates to the 10--20\% level.
Note that once we are successful at such benchmarking experiments,
we will be able to use the ATF
energy spread measurement as a diagnostic for
the absolute emittances of the beam.

\section{CONCLUSION}

We began by describing intrabeam scattering
calculations for electron storage rings, focusing on machines
with small random magnet offset and roll errors.
We derived a simple relation for estimating the relative emittance
growth in $x$ and $y$ due to IBS in such machines.
We have shown that IBS calculations for the ATF
damping ring, when using the formalism of Bjorken-Mtingwa,
a modified formalism of Piwinski (where $\eta^2/\beta$
has been replaced by ${\cal H}$), or a simple high-energy
approximate formula all give results that agree well.
By comparing with numerical results from SAD
we have demonstrated that weak coupling due to random magnet roll
can be approximated by solving the uncoupled problem with the addition
of a global coupling parameter.

Comparing the B-M calculations, and
including the effect of potential well bunch lengthening,
with a complete set of ATF steady-state
energy spread and beam size {\it vs.} current measurements
we have found reasonably good
agreement in
energy spread and horizontal emittance.
At the same time, however,
we find that the measured vertical emittance
is larger than theory in both offset (zero
current emittance) and slope (emittance change with current).
The slope error indicates measurement error and/or
the presence of additional current-dependent physics at the ATF.
The offset error suggests that the assumed Coulomb log is
correct to within a factor of 1.75 (though we believe
that it is, in fact, more accurate, with part of the discrepancy
due to measurement error).
More study is needed.

\begin{acknowledgments}
The authors thank students, staff, and collaborators on the ATF
project. We thank in particular, M.~Ross and A.~Wolski for helpful discussions
on IBS.
One author (K.B.) also
thanks A.~Piwinski for guidance on the topic of
IBS.
\end{acknowledgments}

\bibliography{kbane}

\end{document}